\documentclass{ifacconf}

\usepackage{graphicx}      
\usepackage{natbib}        

\usepackage{graphics} 
\usepackage{times} 
\usepackage{amsmath} 
\usepackage{amssymb}  
\usepackage{algorithm}
\usepackage{xcolor}
\usepackage{arydshln}
\usepackage{microtype}
\usepackage{float}
\usepackage{tikz}
\usepackage{pgfplots}
\pgfplotsset{compat=newest}

\usepackage{bbm}

\newcommand\rmd{\mathrm{d}}

\newcommand\bbr{\mathbb{R}}

\newcommand\calN{\mathcal{N}}
\newcommand\calV{\mathcal{V}}

\usepackage{mleftright}

\usepackage{verbatim}

\usepackage{mathtools,bm}

\usepackage{subfigure}

\newcounter{MYtempeqncnt}
\newlength\fheight
\newlength\fwidth

\begin{document}
\begin{frontmatter}

\title{Data-driven distributed MPC of dynamically coupled linear systems} 

\thanks[footnoteinfo]{%
F. Allg\"ower and M. A. M\"uller are thankful that this work was funded by the Deutsche Forschungsgemeinschaft (DFG, German Research Foundation) - 244600449.
F. Allg\"ower is thankful that this work was funded by the Deutsche Forschungsgemeinschaft (DFG, German Research Foundation) under grant - 468094890; and under Germany’s Excellence Strategy -- EXC 2075 -- 390740016.
M. A. M\"uller is thankful that this project has received funding from the European Research Council
(ERC) under the European Union’s Horizon 2020 research and innovation
programme (grant agreement No 948679).\\
© 2022 the authors. This work has been accepted to IFAC for publication under a Creative Commons Licence CC-BY-NC-ND:
K\"ohler, M., Berberich, J., M\"uller, M.A., and Allg\"ower, F. (2022). Data-driven distributed MPC of dynamically coupled linear systems. \textit{IFAC-PapersOnLine}, 55(30), 365-370.
10.1016/j.ifacol.2022.11.080.
}

\author[First]{Matthias K\"ohler} 
\author[First]{Julian Berberich} 
\author[Second]{Matthias A. M\"uller} 
\author[First]{Frank Allg\"ower}

\address[First]{Matthias K\"ohler (né Hirche), Julian Berberich and Frank Allg\"ower are with the University of Stuttgart, Institute for Systems Theory and Automatic Control, Germany  {\tt\small \{koehler, berberich, allgower\}@ist.uni-stuttgart.de}}
\address[Second]{Matthias A. M\"uller is with the Leibniz University Hannover, Institute of Automatic Control, Germany {\tt\small mueller@irt.uni-hannover.de}}

\begin{abstract}                
    In this paper, we present a data-driven distributed model predictive control (MPC) scheme to stabilise the origin of dynamically coupled discrete-time linear systems subject to decoupled input constraints.
    The local optimisation problems solved by the subsystems rely on a distributed adaptation of the Fundamental Lemma by Willems et al., allowing to parametrise system trajectories using only measured input-output data without explicit model knowledge.
    For the local predictions, the subsystems rely on communicated assumed trajectories of neighbours.
    Each subsystem guarantees a small deviation from these trajectories via a consistency constraint.
    We provide a theoretical analysis of the resulting non-iterative distributed MPC scheme, including proofs of recursive feasibility and (practical) stability.
    Finally, the approach is successfully applied to a numerical example.
\end{abstract}

\begin{keyword}
Data-based control, predictive control, distributed control, large-scale systems, linear systems.
\end{keyword}

\end{frontmatter}
\section{Introduction}
Model predictive control (MPC) is a modern control method with a well-researched theoretical foundation, see e.g.~\cite[]{rawlings2020model,Grune.2017}.
Advantages of MPC are the possibility to explicitly consider constraints on the controlled system and the incorporation of a performance objective.
If the system to be controlled is large-scale or consists of many interconnected but otherwise discernible subsystems, a central controller might not be desirable or even computationally infeasible.
This is alleviated by distributed MPC (DMPC) where a local MPC is designed for each subsystem and the respective control input is computed based on locally available information.
Various different setups for DMPC exist which are suitable for different classes of systems, interconnection structures, and communication topologies using an iterative or non-iterative scheme.
See e.g. the surveys~\cite[]{Christofides.2013, Mueller.2017} or the collection~\cite[]{Maestre.2014} for a selection and classification of various schemes.

Typically, a model derived from first principles is used in MPC to predict the behaviour of the controlled system.
In practice, however, it might be difficult to derive such a model in detail, but obtaining input-output data is comparatively simple.
Recently, for this reason, there has been a push to use only input-output data and no explicit model knowledge to design direct data-driven controllers.
Many results, including our approach, rely on Willems' Fundamental Lemma~\cite[]{willems2005note}, which states that all trajectories of a linear time-varying (LTI) system can be constructed from one persistently exciting trajectory.

Based on this, MPC schemes can be designed using only input-output data and no explicit model knowledge~\cite[]{yang2015data,coulson2019deepc}, even admitting closed-loop guarantees on stability and robustness~\cite[]{berberich2021guarantees}.
Further references may be found in the survey~\cite[]{markovsky2021behavioral}.
Recently, extensions to data-driven distributed MPC for dynamically coupled systems with stability guarantees~\cite[]{allibhoy2021data,alonso2021data} have been made, which require state measurements and state coupling.
Both employ iterative distributed optimisation, requiring a multitude of communication at each time step.

In this paper, we propose a non-iterative data-driven distributed MPC (D3MPC) scheme for dynamically coupled LTI systems.
The local MPC optimisation problems are based on only local input-output data and are solved in parallel assuming that the neighbours follow a previously communicated trajectory.
Therefore, the scheme scales well with the total number of subsystems.
In comparison to the data-driven distributed MPC schemes referenced above, communication is kept to a minimum since in each time step only one trajectory needs to be shared, and a non-iterative parallel implementation is possible, at the price of conservativeness.
We show that the proposed scheme practically asymptotically stabilises the origin of the system
while meeting input constraints.
The main tool is a consistency constraint in the optimisation problems, ensuring that the subsystems' deviation from their communicated trajectories is limited.
The idea is based on~\cite[]{Dunbar.2007} which uses a similar consistency constraint in a model-based setting for continuous-time systems in a stabilising dual-mode DMPC scheme.

\section{Preliminaries}
We denote the natural numbers containing $0$ with $\mathbb{N}_0$.
We write $A \succ 0$ ($A \succcurlyeq 0$) if $A = A^\top$ is positive (semi-)definite.
The smallest eigenvalue of a matrix $A = A^\top$ is denoted by $\lambda_{\min}(A)$. Given another matrix $B = B^\top$, $\lambda_{\min}(A, B) = \min \{\lambda_{\min}(A), \lambda_{\min}(B)\}$.
For a set of vectors $v_1, \dots, v_n$ we denote the stacked vector $v =\begin{bmatrix}
    v_1^\top & \dots v_n^\top
\end{bmatrix}^\top = \mathrm{col}_{i=1}^n(v_i)$.
For a sequence $\{x_k\}_{k=0}^{N-1}$, we define the Hankel matrix
\begin{align*}
    H_L(x)=
    \begin{bmatrix}x_0&x_1&\dots&x_{N-L}\\
    x_1&x_2&\dots&x_{N-L+1}
    \vspace*{-0.4em}
    \\
    \vspace*{-0.4em}
    \vdots&\ddots&\ddots&\vdots\\
    x_{L-1}&x_L&\dots&x_{N-1}
    \end{bmatrix}.
\end{align*}
Moreover, we introduce the notation $x_{[a,b]}= \mathrm{col}_{k=a}^b(x_k)$.
For a (block-)diagonal matrix with $n$ (block-)diagonal elements $A_i$ we write $\mathrm{diag}_{i=1}^n(A_i)$.
The set of integers in the interval $[a, b]$ with $a \le b$ is denoted by $\mathbb{I}_{[a,b]}$.
We make use of class-$\mathcal{K}$, -$\mathcal{K}_\infty$ and -$\mathcal{K}\mathcal{L}$ comparison functions and refer to~\cite[]{Kellett.2014} for their definition.
The cardinality of a set $\mathcal{S}$ is denoted by $\vert \mathcal{S} \vert$.
With $\star$ we denote parts that can be inferred by symmetry.

\begin{defn}
We say that a sequence $\{x_k\}_{k=0}^{N-1}$ with $x_k\in\bbr^n$ is persistently exciting of order $L$ if $\mathrm{rank}(H_L(x))=nL$.
\end{defn}

\section{Data-driven distributed MPC}
\subsection{Distributed system representation}
Our objective is to stabilise a group of $M \ge 2$ dynamically coupled subsystems about the origin.
The dynamic coupling between the subsystems is given by a directed graph $\mathcal{G} = (\mathcal{V}, \mathcal{E})$, where $\mathcal{V} = \{1, \dots, M\}$ is the set of nodes corresponding to subsystems and $\mathcal{E} \subseteq \mathcal{V} \times \mathcal{V}$ is the set of all directed edges between nodes in the graph.
If the output $y^j$ of a subsystem $j$ appears in the dynamic equation of subsystem $i$, and $i \neq j$, then subsystem $j$ is neighbour of subsystem $i$ and part of the set of neighbours $\mathcal{N}_i = \{ j \in \mathcal{V} \mid j \text{ is a neighbour of } i\}$.
Then, $\mathcal{E} = \{ (j, i) \in \mathcal{V} \times \mathcal{V} \mid j \in \mathcal{N}_i \}$.
It is assumed that communication along the directed edges of $\mathcal{G}$ is possible.

We consider as input-output dynamics for subsystem $i\in\calV$:
\begin{subequations}\label{eq:sys_output}
\begin{align}\label{eq:sys_output1}
    x_{t+1}^i &= A_{ii}x_t^i+B_{ii}u_t^i+\sum_{j\in\mathcal{N}_i}B_{ij}y_t^j
    \\
    y_t^i&=C_{ii}x_t^i+D_{ii}u_t^i
    \label{eq:sys_output2}
\end{align}
\end{subequations}
for $i=1,\dots,M$, where $x_t^i\in\mathbb{R}^{n}$ is the state, $u_t^i\in\mathbb{R}^{m}$ is the input, and $y_t^i\in\mathbb{R}^{p}$ is the output of system $i$, all at time $t \in \mathbb{N}_0$.
The input is subject to constraints $u_t^i\in\mathcal{U}_i \subset \mathbb{R}^{m}$ with compact $\mathcal{U}_i$.
We abbreviate $y^{-i} = \mathrm{col}_{j\in\mathcal{N}_i}(y^j)$.
The dimensions of every subsystems' input and output are assumed to be the same, for notational simplicity and without loss of generality.

For each individual subsystem $i$, the matrices $A_{ii}$, $B_{ii}$, $B_{ij}$ with $j\in\mathcal{N}_i$, $C_{ii}$, and $D_{ii}$ are \emph{unknown}, but a data set
$
    \mathcal{D}_i\coloneqq\{\{u_k^{i,\rmd}\}_{k=0}^{N-1},\{y^{i,\rmd}_k\}_{k=0}^{N-1},\{y^{-i,\rmd}_k\}_{k=0}^{N-1}\}
$
that satisfies the dynamics~\eqref{eq:sys_output} is available.

\begin{assum}\label{asm:ctrb_pe_IO}
\begin{enumerate}
\item For each $i \in \mathcal{V}$, the sequence $\begin{bmatrix}u_{[0, N-1]}^{i,\rmd}\\y_{[0, N-1]}^{-i,\rmd}\end{bmatrix}$ is persistently exciting of order $L+n$.

\item $(A, C)$ is observable, where $C = \mathrm{diag}_{i=1}^{M}(C_{ii})$ and
\begin{equation*}
    A = \begin{bmatrix}
        A_{11} & \dots & A_{1M} 
        \vspace*{-0.4em}
        \\
        \vspace*{-0.4em}
        \vdots & \ddots & \vdots \\
        A_{M1} & \dots & A_{MM}
    \end{bmatrix}
\end{equation*}
with $A_{ij} = 0$ if $j \notin \mathcal{N}_i$ and $A_{ij} = B_{ij}C_{jj}$ if $j \in \mathcal{N}_i$.

\item For each $i\in\mathcal{V}$, the pair $(A_{ii}, B_{i})$ is controllable, where 
$
    B_i=
    \begin{bmatrix} 
        B_{ii} & B_{ij_1} & \dots & B_{ij_{|\calN_i|}}
    \end{bmatrix}
$.
\end{enumerate}
\end{assum}

The following lemma is a consequence of Willems' Fundamental Lemma~\cite[]{willems2005note} with the neighbours' outputs interpreted as additional inputs.
It allows the characterisation of any input-output trajectory of~\eqref{eq:sys_output} based on suitable local data $\mathcal{D}_i$
and will be the basis of our D3MPC scheme.
\begin{lem}\label{thm:fundamental_lemma_IO}
Suppose Assumption~\ref{asm:ctrb_pe_IO} holds.
Then, for any $i\in\calV$, $\{u_k^i,y_k^i\}_{k=0}^{L-1}$, $\{y_k^{-i}\}_{k=0}^{L-2}$ is a trajectory of~\eqref{eq:sys_output} if and only if there exists $\alpha^i\in\bbr^{N-L+1}$ such that
\begin{align}\label{eq:wfl}
    \begin{bmatrix}
        H_L(u^{i,\rmd}_{[0, N-1]})
        \\
        H_{L-1}(y^{-i,\rmd}_{[0, N-2]})
        \\
        H_L(y^{i,\rmd}_{[0, N-1]})
    \end{bmatrix}\alpha^i
    =
    \begin{bmatrix}
        u^i_{[0, L-1]}
        \\
        y^{-i}_{[0, L-2]}
        \\
        y^i_{[0, L-1]}
    \end{bmatrix}.
\end{align}
\end{lem}

We only have access to input-output measurements, and not to the minimal state.
Hence, it is useful to define the extended state
\begin{equation}\label{eq:extended_state}
    \xi_t^i = \begin{bmatrix}
        u_{[t-n, t-1]}^i \\ y_{[t-n, t-1]}^{-i} \\ y_{[t-n, t-1]}^i
    \end{bmatrix}.
\end{equation}
With suitable matrices $\tilde{A}_{ii}$, $\tilde{B}_{ii}$, $\tilde{B}_{-i}$, $\tilde{C}_{ii}$ and $\tilde{D}_{ii}$, the system
\begin{subequations}\label{eq:extended_state_system_i}
    \begin{align}
        \xi^i_{t+1} &= \tilde{A}_{ii} \xi^i_t + \tilde{B}_{ii} u_t^i + \tilde{B}_{-i}y_{t}^{-i},
        \label{eq:extended_state_system_i_state}
        \\
        y_t^i &= \tilde{C}_{ii} \xi^i_t + \tilde{D}_{ii}u_t^i,
    \end{align}
\end{subequations}
then has the same input-output behaviour as~\eqref{eq:sys_output}, and there exists $T_{x^i}$ such that $x_t^i = T_{x^i}\xi_t^i$, see, e.g.~\cite[]{goodwin2014adaptive},~\cite[Lemma 2]{koch2021provably}.
In particular,~\eqref{eq:extended_state_system_i_state} looks like~\eqref{eq:extended_state_system_i_state_explicit} with unknown $G_{i,n}, \dots, G_{i,1}$, $S_{i,n}, \dots, S_{i,1}$, $F_{i,n}, \dots, F_{i,1}$ (cf.~\cite[]{berberich2021on}).
Note that~\eqref{eq:wfl} contains only $y_{[0,L-2]}^{-i}$ since there is no feed-through from $y^{-i}$ to $y^i$ in~\eqref{eq:sys_output2}, and thus $y_{L-1}^{-i}$ has no effect on $y_{L-1}^i$.
However, we included $y^{-i}_{t-1}$ in~\eqref{eq:extended_state} since then $\tilde{B}_{-i}$ in~\eqref{eq:extended_state_system_i_state} is known.
We exploit this in the data-driven design of terminal ingredients in Section~\ref{ssec:terminal_ingredients}.
Define also $T_{y^i} = [0 \, \dots \, 0 \, I]$ with $y_t^i = T_{y^i}\xi_{t+1}^i$.
Similarly, a global extended state system 
\begin{subequations}\label{eq:extended_state_global_system}
    \begin{align}
        \xi_{t+1} &= \tilde{A} \xi_t + \tilde{B} u_t
        \\
        y_t &= \tilde{C} \xi_t + \tilde{D}u_t,
    \end{align}
\end{subequations}
can be defined with $\xi_t = \mathrm{col}_{i=1}^M(\xi_t^i)$, $u_t = \mathrm{col}_{i=1}^M(u_t^i)$ and $y_t = \mathrm{col}_{i=1}^M(y_t^i)$ and suitable matrices $\tilde{A}, \tilde{B}, \tilde{C}$ and $\tilde{D}$.

\begin{figure*}[!t]
    \normalsize
    \setcounter{MYtempeqncnt}{\value{equation}}
    \setcounter{equation}{\value{MYtempeqncnt}}
    \begin{align}\label{eq:extended_state_system_i_state_explicit}
        \xi_{t+1}^i =
        \left[\begin{array}{c}
            u^i_{t-n+1} \\ \vdots \\ u^i_{t} \\ \hline
            y^{-i}_{t-n+1} \\ \vdots \\ y^{-i}_{t} \\ \hline
            y^i_{t-n+1} \\ \vdots \\ y^i_{t}
        \end{array}\right]
        =
        \left[\begin{array}{c c c c| c c c c|c c c c}
            0 & I & \dots & 0 & 
            0 & \dots & \dots & 0 & 
            0 & \dots & \dots & 0 
            \\
            \vdots & \ddots & \ddots & \vdots & 
            \vdots & \ddots & \ddots & \vdots & 
            \vdots & \ddots & \ddots & \vdots 
            \\
            0 & \dots & \dots & I & 
            0 & \dots & \dots & 0 & 
            0 & \dots & \dots & 0
            \\
            0 & \dots & \dots & 0 & 
            0 & \dots & \dots & 0 & 
            0 & \dots & \dots & 0
            \\
            \hline
            0 & \dots & \dots & 0 & 
            0 & I & \dots & 0 &  
            0 & \dots & \dots & 0
            \\
            \vdots & \ddots & \ddots & \vdots & 
            \vdots & \ddots & \ddots & \vdots & 
            \vdots & \ddots & \ddots & \vdots 
            \\
            0 & \dots & \dots & 0 & 
            0 & \dots & \dots & I & 
            0 & \dots & \dots & 0
            \\
            0 & \dots & \dots & 0 & 
            0 & \dots & \dots & 0 & 
            0 & \dots & \dots & 0
            \\
            \hline
            0 & \dots & \dots & 0 & 
            0 & \dots & \dots & 0 &  
            0 & I & \dots & 0
            \\
            \vdots & \ddots & \ddots & \vdots & 
            \vdots & \ddots & \ddots & \vdots & 
            \vdots & \ddots & \ddots & \vdots 
            \\
            0 & \dots & \dots & 0 & 
            0 & \dots & \dots & 0 & 
            0 & \dots & \dots & I
            \\
            G_n^i & \dots & \dots & G_1^i & 
            S_n^i & \dots & \dots & S_1^i & 
            F_n^i & \dots & \dots & F_1^i
        \end{array}\right]
        \xi_t^i
        +
        \left[\begin{array}{c}
            0 \\ \vdots \\ I \\ \hline
            0 \\ \vdots \\ 0 \\ \hline
            0 \\ \vdots \\ D_{ii}
        \end{array}\right]
        u_t^i
        +
        \left[\begin{array}{c}
            0 \\ \vdots \\ 0 \\ \hline
            0 \\ \vdots \\ I \\ \hline
            0 \\ \vdots \\ 0
        \end{array}\right]
        y_{t}^{-i}
    \end{align}
    \setcounter{equation}{\value{MYtempeqncnt}+1}
    \hrulefill
    \vspace*{4pt}
    \end{figure*}

\subsection{Data-driven distributed MPC optimisation problem}
Each subsystem locally solves at time step $t$ the following optimisation problem given the measurements $\{u_k,y_k\}_{k=t-n}^{t-1}$ and communicated trajectories $y_{[-n+1, L]}^{-i,*}(t-1)$ of the neighbours.
\vspace*{-0.5em}
\begin{subequations}\label{eq:D3MPC_IO}
\begin{align}
    &\min_{\alpha^i(t)} \sum_{k=0}^{L-1}\lVert y_k^i(t)\rVert_{Q_i}^2
    +\lVert u_k^i(t)\rVert_{R_i}^2
    + \Vert \xi_L^i(t) \Vert_{P_i}^2
    \intertext{subject to}
    &
    \label{eq:D3MPC_IO_hankel}
    \begin{bmatrix}
        H_{L+n}(u_{[0,N-1]}^{i,\rmd})
        \\
        H_{L+n-1}(y_{[0,N-2]}^{-i,\rmd})
        \\
        H_{L+n}(y_{[0,N-1]}^{i,\rmd})
    \end{bmatrix}\alpha^i(t)
    =
    \begin{bmatrix}
        u_{[-n, L-1]}^i(t)
        \\
        y^{-i,*}_{[-n+1,L-1]}(t-1)
        \\
        y_{[-n, L-1]}^i(t)
    \end{bmatrix}
    \\
    \label{eq:D3MPC_IO_init}
    & \begin{bmatrix}u_{[-n,-1]}^i(t)\\y_{[-n,-1]}^i(t)\end{bmatrix}
    =\begin{bmatrix}u_{[t-n,t-1]}^i\\y_{[t-n,t-1]}^i\end{bmatrix}
    \\
    \label{eq:D3MPC_IO_input}
    & u^{i}_k(t) \in \mathcal{U}_i, \; k \in \mathbb{I}_{[0,L-1]},
    \\
    \label{eq:D3MPC_IO_terminal}
    & \xi_L^i(t)\in \mathcal{X}_i^\mathrm{f}(\theta_i \epsilon_i),
    \\
    \label{eq:D3MPC_IO_consistency_input}
    & \Vert u_k^i(t) - {u}_{k+1}^{i,*}(t-1) \Vert^2 \le \Vert \hat{u}_k^i(t) - {u}_{k+1}^{i,*}(t-1) \Vert^2 + \Omega_i,
    \notag \\
    & k \in \mathbb{I}_{[0,L-1]},
    \\
    \label{eq:D3MPC_IO_consistency_output}
    & \Vert y_k^i(t) - {y}_{k+1}^{i,*}(t-1) \Vert^2 \le \Vert \hat{y}_k^i(t) - {y}_{k+1}^{i,*}(t-1) \Vert^2 + \Omega_i,
    \notag \\
    & k \in \mathbb{I}_{[0,L-1]},
    \\
    \label{eq:D3MPC_IO_extended_state}
    &\xi_L^i(t) =
    \begin{bmatrix}
        u_{[L-n,L-1]}^i(t)
        \\
        y_{[L-n+1,L]}^{-i,*}(t-1)
        \\
        y_{[L-n,L-1]}^i(t)
    \end{bmatrix},
\end{align}
\end{subequations}
with a constant $\Omega_i \ge 0$, and momentarily defined trajectories $\hat{u}_{[0,L-1]}^i(t)$ and $\hat{y}_{[0,L-1]}^i(t)$ as well as a terminal cost matrix $P_i$ and set $\mathcal{X}_i^\mathrm{f}$.
The optimal solution of~\eqref{eq:D3MPC_IO} is denoted by $\alpha^{i,*}(t)$ with corresponding optimal input sequence  $u^{i,*}_{[-n,L-1]}(t)$ and predicted output sequence $y^{i,*}_{[-n,L-1]}(t)$.
Constraint~\eqref{eq:D3MPC_IO_hankel} is based on Lemma~\ref{thm:fundamental_lemma_IO} and used to predict the input-output behaviour of the system over the prediction horizon $L$, whereas~\eqref{eq:D3MPC_IO_init} fixes the initial condition.
The predicted extended state~\eqref{eq:D3MPC_IO_extended_state} is confined to a terminal set through~\eqref{eq:D3MPC_IO_terminal} which is tightened using a factor $\theta_i > 0$ to provide recursive feasibility.
Each subsystem assumes that the outputs of their neighbours follow a communicated trajectory (cf.~\eqref{eq:D3MPC_IO_hankel}).
The neighbours, however, will in general deviate from this trajectory. 
To ensure recursive feasibility despite this discrepancy, the consistency constraints~\eqref{eq:D3MPC_IO_consistency_input} and~\eqref{eq:D3MPC_IO_consistency_output} are included.
It forces the predicted input and output to stay as close to what has been communicated as the in~\ref{ssec:d3mpc_scheme} defined feasible trajectories $\hat{u}^i_{[0,L-1]}(t)$ and $\hat{y}^i_{[0,L-1]}(t)$ can.
If $\mathcal{U}_i$ and $\mathcal{X}_i^\mathrm{f}(\theta_i\epsilon_i)$ are ellipsoidal or polytopic,~\eqref{eq:D3MPC_IO} is a convex quadratically constraint quadratic program, which can be solved efficiently.

\subsection{Terminal ingredients}\label{ssec:terminal_ingredients}
\begin{assum}\label{asm:terminal_ingredients_data_IO}
    For every subsystem $i$, there exist $P_i \succ 0$, $\mu_i, \eta_i,\epsilon_i > 0$, $\theta_i \in (0, 1)$ and a terminal controller $\kappa_i$ such that
    \begin{align}
        \hspace{-0.4em}
        \Vert \xi^{i,+} \Vert_{P_i}^2 {- \Vert \xi^{i} \Vert_{P_i}^2}
         &\le - \eta_i\Vert \xi^i \Vert^2 {- \Vert y^i \Vert_{Q_i}^2} {- \Vert \kappa_i(\xi^i) \Vert_{R_i}^2} + \mu_i \label{eq:terminal_cost_decrease_data_IO}
        \\
        \kappa_i(\xi^i) &\in \mathcal{U}_i
        \label{eq:terminal_controller_feasible_data_IO}
        \\
        \Vert \xi^{i,+} \Vert_{P_i}^2 & \le \theta_i \epsilon_i
        \label{eq:terminal_contraction}
    \end{align}
    if $\xi^i \in \mathcal{X}_i^\mathrm{f}(\epsilon_i) = \{\xi^i \mid \Vert \xi^i \Vert_{P_i}^2 \le \epsilon_i \}$ for all $i \in \mathcal{V}$ and where $\xi^{i,+} = \tilde{A}_{ii}\xi^i + \sum_{j\in\mathcal{N}_i} \tilde{A}_{ij} y^j + \tilde{B}_{ii}\kappa_i(\xi^i)$.
\end{assum}
This assumption implies that inside ${X}_i^\mathrm{f}(\epsilon_i)$ every subsystem can be driven into an invariant set strictly inside ${X}_i^\mathrm{f}(\epsilon_i)$ despite the influence of the neighbours' outputs.
Implicitly, the dynamic coupling needs to be sufficiently weak.
If $\mu_i$ is sufficiently small, it can be shown that~\eqref{eq:terminal_cost_decrease_data_IO} implies~\eqref{eq:terminal_contraction} if $\xi^i \in \mathcal{X}_i^\mathrm{f}(\epsilon_i)$ for all $i \in \mathcal{V}$.

We comment on how the method in~\cite[]{berberich2021on} could be adapted to compute distributed terminal ingredients as in Assumption~\ref{asm:terminal_ingredients_data_IO} from input-output data, if the coupling is sufficiently weak. 
A general data-driven method to compute these terminal ingredients is left open for future research.
We rewrite~\eqref{eq:extended_state_system_i} into
\vspace*{-1em}
\begin{subequations}
\begin{align*}
    \left[\begin{array}{c}
        \xi^i_{t+1} \\
        \hline
        z_t^i
    \end{array}\right]
    =
    \left[\begin{array}{c | c c c}
        \bar{A}_i & B_{u^i} & B_{y^{-i}} & B_{w^i}\\
        \hline
        \begin{bmatrix} I \\ 0 \end{bmatrix}
        &
        \begin{bmatrix} 0 \\ I \end{bmatrix}
        &
        0
        &
        0
    \end{array}\right]
    \left[\begin{array}{c}
        \xi^i_{t} \\
        \hline
        u_t^i \\
        y_t^{-i} \\
        w_t^i 
    \end{array}\right],    
\end{align*}
\end{subequations}
where $w_t^i = \Delta_i z_t^i$ contains all unknown elements in~\eqref{eq:extended_state_system_i}, i.e. $\Delta_i = [G_{i,n}, \dots, G_{i,1}, S_{i,n}, \dots, S_{i,1}, F_{i,n}, \dots, F_{i,1}, D_{ii}]$.
Define the matrices
\begin{align*}
    \Xi_i &= \left[\xi_n^{\mathrm{d},i} \, \xi_{n+1}^{\mathrm{d},i} \, \dots \, \xi_{N-1}^{\mathrm{d},i} \right], 
    \, 
    \Xi_i^+ = \left[ \xi_{n+1}^{\mathrm{d},i} \, \xi_{n+2}^{\mathrm{d},i} \, \dots \, \xi_{N}^{\mathrm{d},i} \right]
    \\
    U_i &= \left[u_n^{\mathrm{d},i} \, u_{n+1}^{\mathrm{d},i} \, \dots \, u_{N-1}^{\mathrm{d},i} \right],  Y_{-i} = \left[ y_n^{\mathrm{d},-i} \, y_{n+1}^{\mathrm{d},-i} \, \dots \, y_{N-1}^{\mathrm{d},-i} \right], 
\end{align*}
and $Z_i = \begin{bmatrix}\Xi_i^\top & U_i^\top\end{bmatrix}^\top$ with the extended state $\xi_t^{\mathrm{d},i}$ based on the available input-output data $\mathcal{D}_i$.
Further, define $M_i = \Xi^+_i - \bar{A}_i \Xi_i - B_{u^i} U_i - B_{y^{-i}}Y_{-i}$ and the matrix
\begin{equation*}
    \bar{P}_{\Delta_i}^{w^i} = \begin{bmatrix}
        \star
    \end{bmatrix}^\top
    \begin{bmatrix}
        -Z_iZ_i^{\top} & Z_i M_i^\top B_{w^i} \\
        B_{w^i}^\top M_iZ_i^\top & -B_{w^i}^\top M_i M_i^\top B_{w^i}   
    \end{bmatrix}
    \begin{bmatrix}
        0 & I \\ B_{w^i}^\top & 0
    \end{bmatrix}.
\end{equation*}
\begin{lem}[cf.~{\cite[Proposition 10]{berberich2021on}}]\label{lm:terminal_ingredients_lmi}
    \hfill\\
    Factorise $T_{y^i}^\top Q_i T_{y^i} = Q_{i,\mathrm{r}}^\top Q_{i,\mathrm{r}}$ and $R_i = R_{i,\mathrm{r}}^\top R_{i,\mathrm{r}}$.
    Suppose there exist $\mathcal{X}_i \succ 0$, $\Gamma_i \succ 0$, $\mathcal{M}_i$, $\tau_i \ge 0$, $\gamma_i > 0$ such that $\mathrm{trace}(\Gamma_i) < \gamma_i^2$, $\begin{bmatrix}
        \Gamma_i & I \\ I & \mathcal{X}
    \end{bmatrix} \succ 0$, and
    \begin{equation*}
        \begin{bmatrix}
        \tau \bar{P}_{\Delta_i}^{w^i} - \begin{bmatrix}
            \mathcal{X}_i & 0 \\ 0 & 0
        \end{bmatrix}
        &
        \begin{bmatrix}
            \bar{A}^i\mathcal{X}_i + B_{u^i} \mathcal{M}_i \\ \mathcal{X}_i \\ \mathcal{M}_i
        \end{bmatrix}
        &
        0
        \\
        \star & -\mathcal{X}_i & \begin{bmatrix}
            Q_{i,\mathrm{r}}\mathcal{X}_i \\ R_{i,\mathrm{r}}\mathcal{M}_i
        \end{bmatrix}^\top
        \\
        \star & \star & -I
        \end{bmatrix} \prec 0.
    \end{equation*}
    Define $P_i = \mathcal{X}^{-1}_i - T_{y^i}^\top Q_i T_{y^i}$ and $K_i = \mathcal{M}_i\mathcal{X}_i^{-i}$.
    Then, there exists $\bar{\eta}_i > 0$ such that
    \begin{equation}\label{eq:decrease_wo_neighbour}
        \Vert \bar{\xi}^{i,+} \Vert_{P_i}^2 - \Vert \xi^i \Vert_{P_i}^2 
        \le
        -\bar{\eta}_i \Vert \xi^i \Vert_{P_i}^2 
        - \Vert y^i \Vert_{Q_i}^2
        - \Vert \kappa_i(\xi^i) \Vert_{R_i}^2 
    \end{equation}
    with $\bar{\xi}^{i,+} = (\tilde{A}_{ii} + \tilde{B}_{ii}K_i) \xi^i$ and $\kappa_i(\xi^i) = K_i\xi^i$.
\end{lem}
This result is a slight extension of~\cite[Proposition 10]{berberich2021on} and we refer to~\cite[]{berberich2021on} for the proof.
The existence of the additional parameter $\bar{\eta}_i$ is guaranteed by strictness of the LMIs in~\cite[Proposition 10]{berberich2021on}.
Note that~\eqref{eq:decrease_wo_neighbour} is similar to~\eqref{eq:terminal_cost_decrease_data_IO}, except that no coupling is considered, cf. $\bar{\xi}^{i,+}$ in~\eqref{eq:decrease_wo_neighbour} to $\xi^{i,+}$ in~\eqref{eq:terminal_cost_decrease_data_IO}.
Thus, the neglected dynamic coupling needs to be sufficiently weak for~\eqref{eq:decrease_wo_neighbour} to imply~\eqref{eq:terminal_cost_decrease_data_IO}.
Furthermore, if a bound on the neighbours' outputs is known, they can be interpreted as bounded noise and it is straightforward to adapt the method in~\cite[]{berberich2021on} to design a robust feedback law, which would alleviate this issue.
Feasibility of the conditions in Lemma~\ref{lm:terminal_ingredients_lmi} requires a potentially restrictive condition on the dimensions of subsystem~\eqref{eq:sys_output}, as further discussed in~\cite[]{berberich2021on}.

\subsection{Distributed data-driven MPC scheme}\label{ssec:d3mpc_scheme}
The following assumption bypasses the difficult task of constructing an initially feasible candidate in~\eqref{eq:D3MPC_IO}.
\begin{assum}\label{asm:feasible_initial_trajectory}
    At time $t = 0$, for all subsystems $i \in \mathcal{V}$, there exists $\hat{\alpha}^i(0)$ with corresponding $\hat{y}^i_{[-n,L-1]}(0)$ such that it is a feasible candidate in~\eqref{eq:D3MPC_IO} for $y^{-i,*}_{[-n+1, L]}(-1) = \hat{y}^{-i}_{[-n,L-1]}(0)$.
    In addition, each subsystem knows $\hat{\alpha}^i(0)$.
\end{assum}
Since the prediction relies on communicated trajectories of the neighbours, the first communicated output trajectory is $\hat{y}^i_{[-n,L-1]}(0)$ from Assumption~\ref{asm:feasible_initial_trajectory}.

After solving~\eqref{eq:D3MPC_IO} at $t-1$, the predicted output trajectory of subsystem $i$, $y_{[0,L-1]}^{i,*}(t-1)$, is available for communication after a one-step extension as described momentarily.
However, in general, it does not correspond to a feasible candidate at time $t$ because of~\eqref{eq:D3MPC_IO_hankel}, which now depends on the updated $y_{[-n+1,L]}^{-i,*}(t-1)$, whereas $y_{[0,L-1]}^{i,*}(t-1)$ was computed based on $y_{[-n+1,L]}^{-i,*}(t-2)$.
Instead, a feasible candidate at time $t \in \mathbb{N}$ can be constructed using
\begin{subequations}\label{eq:input_candidate}
    \begin{align}
        \hat{u}^i_{[-n,L-2]}(t) &= u_{[-n+1,L-1]}^{i,*}(t-1),
        \\
        \hat{u}^i_{L-1}(t) &= \kappa_i(\hat{\xi}_{L-1}^i(t)), \label{eq:input_candidate_b}
        \\
        \label{eq:candidate_extended_state_L-1}
        \hat{\xi}^i_{L-1}(t) &=
        \begin{bmatrix}
            \hat{u}_{[L-n-1,L-2]}^i(t)  \\
            y_{[L-n,L-1]}^{-i,*}(t-1)  \\
            \hat{y}_{[L-n-1,L-2]}^i(t)
        \end{bmatrix},
    \end{align}
\end{subequations}
as will be shown below.
The corresponding candidate $\hat{\alpha}^i(t)$ and output trajectory $\hat{y}^i_{[-n,L-1]}(t)$ are computed with Algorithm~\ref{alg:DD_sim} based on the updated neighbours' trajectories $y_{[-n+1,L]}^{-i,*}(t-1)$ and output measurement $y^i_{[t-n, t-1]} = y_{[-n+1,0]}^{i,*}(t-1)$.
\begin{algorithm}[t]
    \begin{alg}\label{alg:DD_sim}
    \normalfont{\textbf{Data-driven simulation (subsystem $i$)}} (cf.~\cite[]{markovsky2008data})\\
    \textbf{Input:} 
    \begin{itemize}
    \vspace*{-1em}
    \item Data $\mathcal{D}_i$, where 
    $\begin{bmatrix}
        u^\rmd_{[0,N-1]}
        \\
        y^{\rmd,-i}_{[0,N-2]}
    \end{bmatrix}$ 
    is persistently exciting of order $L+2n$ and $(A_{ii}, B_i)$ is controllable.
    
    \item Initial condition $\{u_k^i,y_k^i,y_k^{-i}\}_{k=-n}^{-1}$.
    
    \item New input and neighbours' output data $\{u_k^i\}_{k=0}^{L-1}$, $\{y_k^{-i}\}_{k=0}^{L-2}$.
    \end{itemize}
    \textbf{Procedure:}
    \begin{enumerate}
    \item Compute $\alpha^i$ satisfying
    \begin{align}\label{eq:data_driven_sim}
        \begin{bmatrix}
            H_{L+n}(u^{\rmd,i}_{[0, N-1]})
            \\
            H_{L+n-1}(y^{\rmd,-i}_{[0, N-2]})
            \\
            H_{n}(y^{\rmd,i}_{[0, N-L-1]})
        \end{bmatrix}
        \alpha^i
        =
        \begin{bmatrix}
            u^i_{[-n,L-1]}
            \\
            y^{-i}_{[-n,L-2]}
            \\
            y^i_{[-n,-1]}
        \end{bmatrix}
    \end{align}
    
    \item Compute $y_{[0,L-1]}^i=H_{L}(y^{\rmd,i}_{[n,N-1]})\alpha^i$.
    \end{enumerate}
    \textbf{Output:}
    Resulting simulated output trajectory $\{y_k\}_{k=0}^{L-1}$.
    \end{alg}
\end{algorithm}

The D3MPC scheme is stated in Algorithm~\ref{alg:D3MPC_scheme}.
\begin{algorithm}[t]
\begin{alg}\label{alg:D3MPC_scheme}
\normalfont{\textbf{Data-driven distributed MPC scheme}}\\
\textbf{Input for all $i \in \mathcal{V}$:} 
\begin{itemize}
    \vspace*{-0.6em}
    \item Data $\mathcal{D}_i$, where 
    $
    \begin{bmatrix}
        u^\rmd
        \\
        y^{\rmd,-i}
    \end{bmatrix}
    $ is persistently exciting of order $L+2n$ and $(A_{ii}, B_i)$ is controllable.
    \item Initial measurement $\{ u_k^i, y_k^i \}_{k=-n}^{-1}$.
    \item Initially communicated trajectories $y^{-i,*}_{[-n+1, L]}(-1)$.
\end{itemize}
\textbf{Procedure:}
For all $i \in \mathcal{V}$:
\begin{enumerate}
    \item Solve the local MPC problem~\eqref{eq:D3MPC_IO}.
    \item Apply $u_t = u_0^{i,*}(t)$ and measure $y_{t}^i$.
    \item Compute $u_L^{i,*}(t) = \kappa_i(\xi^{i,*}_L(t))$. Then, compute $y_L^{i,*}(t)$ using Algorithm~\ref{alg:DD_sim} with initial condition
    ${u}_{[-n,-1]}^{i,*}(t)$, $y_{[-n+1,0]}^{-i,*}(t-1)$, $y_{[-n,-1]}^{i,*}(t)$ and new input and neighbours' output trajectories
    ${u}^{i,*}_{[0,L]}(t)$, $y_{[1,L]}^{-i,*}(t-1)$.
    \item Send $y_{[-n+1,L]}^{i,*}(t)$; receive $y_{[-n+1,L]}^{-i,*}(t)$ from neighbours.
    \item Set $t = t + 1$.
    \item Compute $\hat{u}_{[-n,L-1]}^i(t)$ according to~\eqref{eq:input_candidate} as well as $\hat{y}_{[-n,L-1]}^i(t)$ using Algorithm~\ref{alg:DD_sim} with initial condition
    $\hat{u}^i_{[-n,-1]}(t)$, $y_{[-n+1,0]}^{-i,*}(t-1)$, $y_{[-n,-1]}^{i,*}(t)$ and new input and neighbours' output trajectories
    $\hat{u}^i_{[0,L-1]}(t)$, $y_{[1,L-1]}^{-i,*}(t-1)$.
\end{enumerate}
\end{alg}
\end{algorithm}
Note that each subsystem can solve its MPC problem in parallel in Step 1) of Algorithm~\ref{alg:D3MPC_scheme}.
Hence, the complexity of the scheme increases only with the number of neighbours each subsystem has, but not with the total number of subsystems.
In addition, since communication is necessary only once in each time step, the communication overhead is kept to a minimum.
In Step 4) of Algorithm~\ref{alg:D3MPC_scheme}, we compute an extension of the optimal output trajectory by one step, which is denoted by $y_L^{i,*}(t)$ with a slight abuse of notation.

\section{Closed-loop guarantees}
In this section, we prove that the origin is practically asymptotically stabilised if Algorithm~\ref{alg:D3MPC_scheme} is applied to the system~\eqref{eq:extended_state_global_system}.
An important requirement to this end is that in each time step each subsystem is able to solve the MPC optimisation problem~\eqref{eq:D3MPC_IO}, i.e.~\eqref{eq:D3MPC_IO} is recursively feasible, which in particular implies that the input constraints are not violated.

\subsection{Recursive feasibility}\label{ssec:feasibility}
The following assumption is crucial to prove recursive feasibility of~\eqref{eq:D3MPC_IO} and captures the central idea of the scheme.
If all subsystems stay close to their communicated trajectory, the unexpected influence on neighbours is sufficiently bounded.
\begin{assum}\label{asm:extended_state_bound}
    There exist $\sigma_i^\prime, \tilde{\sigma}_i \in \mathcal{K}$ such that
    $ \Vert \hat{\xi}^i_{L-1}(t+1) - \xi^{i,*}_L(t) \Vert_{P_i} \le \tilde{\sigma}_i(\Omega_i)$
    and 
    $\Vert \hat{y}^i_{k}(t+1) - y^{i,*}_{k+1}(t) \Vert_{Q_i} \le \sigma_i^\prime(\Omega_i)$
    for all $t \in \mathbb{N}_0$ and $k \in \mathbb{I}_{[0,L-2]}$ if Algorithm~\ref{alg:D3MPC_scheme} is used, with $\hat{\xi}_{L-1}^i(t+1)$ as in~\eqref{eq:candidate_extended_state_L-1} and where
    $
        \xi^{i,*}_L(t) =
        \begin{bmatrix}
            u_{[L-n,L-1]}^{i,*}(t)^\top &
            y_{[L-n+1,L]}^{-i,*}(t-1)^\top  &
            y_{[L-n,L-1]}^{i,*}(t)^\top
        \end{bmatrix}^\top.
    $
\end{assum}
As stated in~\cite[]{Dunbar.2007}, it is to be expected that this requires sufficiently weak dynamic coupling.
Although we conjecture that Assumption~\ref{asm:extended_state_bound} can be shown to hold (cf.~\cite[Lemma 3]{Dunbar.2007}), it is not straightforward to compute $\tilde{\sigma}_i$ and $\sigma_i^\prime$ only based on the available input-output data.
This is beyond the scope of this paper, yet we want to highlight two possibilities.
It may be possible to extract this information from the Hankel matrix of each subsystem in Lemma~\ref{thm:fundamental_lemma_IO}, since it contains the dynamic coupling.
Alternatively, one may be able to use bounds on the norms of $A_{ii}$ and $B_{ij}$, $j \in \mathcal{N}_i$, in~\eqref{eq:sys_output} to construct $\tilde{\sigma}_i$ and $\sigma_i^\prime$.
Using a similar approach as in~\cite[]{Wildhagen.04012021}, the latter can be estimated using input-output data.

We now show that $\hat{u}_{[0,L-1]}(t)$ from~\eqref{eq:input_candidate} leads to a feasible candidate in~\eqref{eq:D3MPC_IO} and the MPC problem is recursively feasible.
\begin{thm}\label{thm:recursive_feasibility}
    Let Assumptions~\ref{asm:ctrb_pe_IO}--\ref{asm:extended_state_bound} hold.
    In particular, from Assumption~\ref{asm:feasible_initial_trajectory}, let the MPC problem~\eqref{eq:D3MPC_IO} be feasible for all $i \in \mathcal{V}$ at time $t = 0$.
    Then,~\eqref{eq:D3MPC_IO} is also feasible for all $i \in \mathcal{V}$ and all $t \in \mathbb{N}$, 
    if
    $\tilde{\sigma}_i(\Omega_i) \le (1 - \sqrt{\theta_i})\sqrt{\epsilon_i}$
    and 
    $\theta_i \ge 1 - \eta_i \lambda_{\max}(P_i)^{-1}$ for all $i \in \mathcal{V}$ with $\epsilon_i$, $\eta_i$, and $P_i$ from Assumption~\ref{asm:terminal_ingredients_data_IO}.
\end{thm}
\begin{pf}
    Let $i \in \mathcal{V}$. 
    From Assumption~\ref{asm:feasible_initial_trajectory}, a solution $\alpha^{i,*}(0)$ with corresponding $u_{[-n,L-1]}^{i,*}(0)$ and $y_{[-n,L-1]}^{i,*}(0)$ to~\eqref{eq:D3MPC_IO} exists.
    Assume for induction that~\eqref{eq:D3MPC_IO} is feasible at time $t$.
    Consider the candidate input trajectory $\hat{u}_{[-n,L-1]}^{i}(t+1)$ from~\eqref{eq:input_candidate} together with the updated output data of the neighbours $y^{-i,*}_{[-n+1,L]}(t)$ which, if applied in~\eqref{eq:sys_output1}, yields $\hat{y}_{[-n,L-1]}^{i}(t+1)$.
    By Lemma~\ref{thm:fundamental_lemma_IO}, there exists a corresponding $\hat{\alpha}^{i}(t+1)$ such that \eqref{eq:D3MPC_IO_hankel} is satisfied.
    By definition of $\hat{u}_{[-n,L-1]}^{i}(t+1)$ and $\hat{y}_{[-n,L-1]}^{i}(t+1)$ the constraint \eqref{eq:D3MPC_IO_init} holds.
    From feasibility of $u^{i,*}_{[-n,L-1]}(t)$ it follows that $\hat{u}_{[-n,L-2]}^i(t+1) \in \mathcal{U}_i$.
    Clearly~\eqref{eq:D3MPC_IO_consistency_input} and~\eqref{eq:D3MPC_IO_consistency_output} hold.
    It remains to be shown that~\eqref{eq:D3MPC_IO_input} for $k = L-1$ and~\eqref{eq:D3MPC_IO_terminal} are satisfied.
    Feasibility of~\eqref{eq:D3MPC_IO} at time $t$ and Assumption~\ref{asm:extended_state_bound} yield 
    $
        \Vert \hat{\xi}^i_{L-1}(t+1) \Vert_{P_i} 
        \le 
        \Vert {\xi}^{i,*}_{L}(t) \Vert_{P_i} + \Vert \hat{\xi}^i_{L-1}(t+1) - {\xi}^{i,*}_{L}(t) \Vert_{P_i}
        \le 
        \sqrt{\theta_i\epsilon_i }+ \tilde{\sigma}_i(\Omega_i) \le \sqrt{\epsilon_i},
    $
    since $\tilde{\sigma}_i(\Omega_i) \le (1 - \sqrt{\theta_i})\sqrt{\epsilon_i}$ with $\theta_i \in (0, 1)$.
    Hence, $\hat{\xi}^i_{L-1}(t+1) \in \mathcal{X}_i^\mathrm{f}(\epsilon_i)$ and by Assumption~\ref{asm:terminal_ingredients_data_IO}, \eqref{eq:D3MPC_IO_input} is satisfied for $k = L-1$ and for all $i \in \mathcal{V}$.
    In addition, from~\eqref{eq:terminal_contraction} in Assumption~\ref{asm:terminal_ingredients_data_IO},
    $\hat{\xi}^i_{L}(t+1) \in \mathcal{X}_i^\mathrm{f}(\theta_i \epsilon_i)$ for all $i \in \mathcal{V}$.
    \hspace{\fill} \qed
\end{pf}

\subsection{Practical stability}\label{ssec:stability}
We define
$
    V_t^* = \sum_{i=1}^M \sum_{k=0}^{L-1} (\Vert y_k^{i,*}(t) \Vert_{Q_i}^2 + \Vert u_k^{i,*}(t) \Vert_{R_i}^2 + \Vert \xi^{i,*}_L(t) \Vert_{P_i}^2)
$
and assume the following.
\begin{assum}\label{asm:V_t_upper_bound}
    There exists $c_{\mathrm{ub}} > 0$ such that
    $V_t^* \le c_{\mathrm{ub}} \Vert \xi_t \Vert^2$
    holds for all $t \in \mathbb{N}_0$.
\end{assum}
This assumption can be shown to hold if, e.g. $\mathcal{U}_i$ are compact polytopes for all $i \in \mathcal{V}$ and the cost of the initially feasible candidate from Assumption~\ref{asm:feasible_initial_trajectory} admits a quadratic upper bound.

The following theorem establishes practical stability of the origin of the global extended system~\eqref{eq:extended_state_global_system}. 
Hence, also of the closed-loop state, since $x_t^i = T_{x^i}\xi_t^i$ for all $i \in \mathcal{V}$.
It shows a trade-off between giving each subsystem a larger margin of freedom ($\Omega_i$ larger), compared to tight stabilisation of the origin ($\Omega_i$ smaller).
\begin{thm}\label{thm:stability}
    Let Assumptions~\ref{asm:ctrb_pe_IO}--\ref{asm:extended_state_bound} hold.
    If the D3MPC scheme in Algorithm~\ref{alg:D3MPC_scheme} is applied, then the origin of the resulting closed-loop system is practically stable.
    That is, there exist $\beta \in \mathcal{KL}$, $\delta_1, \delta_2 \in \mathcal{K}_\infty$ and $\Omega_{\max}$ such that for all $\Omega \leq \Omega_{\max}$ 
    \begin{equation*}
        \Vert \xi_t \Vert \le \beta(\Vert \xi_0 \Vert, t) + \delta_1(\Omega) + \delta_2(\mu)
    \end{equation*}
    holds for the closed-loop solution of~\eqref{eq:extended_state_global_system}, where $\Omega = \max_i \Omega_i$ and $\mu = \sum_{i=1}^m \mu_i$.
\end{thm}
\begin{pf}
    For brevity, we write $\hat{\cdot} \coloneqq \hat{\cdot}(t+1)$, e.g. $\hat{\xi}^{i}_{L} \coloneqq \hat{\xi}^{i}_{L}(t+1)$, and ${\cdot}^* \coloneqq {\cdot}^*(t)$, e.g. ${y}_{k+1}^{i,*} \coloneqq {y}_{k+1}^{i,*}(t)$.
    Since $\hat{\alpha}^{i}(t+1)$ with $\hat{u}_{[-n,L-1]}^{i}(t+1)$ and $\hat{y}_{[-n,L-1]}^{i}(t+1)$ is a feasible choice in~\eqref{eq:D3MPC_IO} at time $t+1$ (see Theorem~\ref{thm:recursive_feasibility}) and $\hat{u}_k^i(t+1) = u_{k+1}^{i,*}(t)$ for $k \in \mathbb{I}_{[0,L-2]}$, 
    $
        V_{t+1}^* \le \sum_{i=1}^M ( \sum_{k=0}^{L-1} \Vert \hat{y}_k^{i} \Vert_{Q_i}^2 + \Vert \hat{u}_k^{i} \Vert_{R_i}^2  + \Vert \hat{\xi}^{i}_L \Vert_{P_i}^2 )
        = V_t^* 
        + \sum_{i=1}^M 
            (
                \sum_{k=0}^{L-2} \Vert \hat{y}_k^{i} \Vert_{Q_i}^2 - \Vert {y}_{k+1}^{i,*} \Vert_{Q_i}^2
            ) 
        + \sum_{i=0}^M
            (
                \Vert \hat{y}^i_{L-1} \Vert_{Q_i}^2 + \Vert \hat{u}^i_{L-1} \Vert_{R_i}^2 + \Vert \hat{\xi}^{i}_L \Vert_{P_i}^2 - \Vert \hat{\xi}^{i}_{L-1} \Vert_{P_i}^2 
                + \Vert \hat{\xi}^{i}_{L-1} \Vert_{P_i}^2 - \Vert {\xi}^{i,*}_L \Vert_{P_i}^2 - \Vert y_t^{i} \Vert_{Q_i}^2 - \Vert u_t^{i} \Vert_{R_i}^2
            )
        .
    $
    Then, since $\hat{\xi}^i_{L-1}(t+1) \in  \mathcal{X}_i^\mathrm{f}(\epsilon_i)$ for all $i \in \mathcal{V}$ as shown in the proof of Theorem~\ref{thm:recursive_feasibility}, and by definition of $\hat{u}^i_{L-1}(t+1)$ in~\eqref{eq:input_candidate_b}, from Assumption~\ref{asm:terminal_ingredients_data_IO} it follows that
    $
        V_{t+1}^* - V_t^* \le - \Vert y_t \Vert_{Q}^2 - \Vert u_t \Vert_{R}^2 
        + {\sum_{i=1}^M} 
        (
            {\sum_{k=0}^{L-2}}
            \Vert \hat{y}_k^{i} \Vert_{Q_i}^2 - \Vert {y}_{k+1}^{i,*} \Vert_{Q_i}^2
        )
        + \sum_{i=0}^M (\Vert \hat{\xi}^{i}_{L-1} \Vert_{P_i}^2 - \Vert {\xi}^{i,*}_L \Vert_{P_i}^2
        + \mu_i),
    $
    with $Q = \mathrm{diag}_{i\in\mathcal{V}}(Q_i)$ and $R = \mathrm{diag}_{i\in\mathcal{V}}(R_i)$.
    Note that for vectors $a, b$, $\Vert a \Vert^2 - \Vert b \Vert^2 \le \Vert a-b \Vert^2 + 2 \Vert b \Vert \Vert a-b \Vert$.
    Hence, with $\Vert {y}_{k+1}^{i,*}(t) \Vert_{Q_i}^2 \le V_t^*$ for all $k \in \mathbb{I}_{[0:L-2]}$, and $\Vert {\xi}^{i,*}_L(t) \Vert_{P_i}^2 \le V_t^*$ for all $i \in \mathcal{V}$,
    $
        V_{t+1}^* - V_t^* \le - \Vert y_t \Vert_{Q}^2 - \Vert u_t \Vert_{R}^2 + \mu
        + {\sum_{i=1}^M}
        (
            {\sum_{k=0}^{L-2}} 
            \Vert \hat{y}_k^{i} - {y}_{k+1}^{i,*} \Vert_{Q_i}^2 + 2\sqrt{V_t^*} \Vert \hat{y}_k^{i} - {y}_{k+1}^{i,*} \Vert_{Q_i}
        ) 
        + {\sum_{i=0}^M}
        (
            \Vert \hat{\xi}^{i}_{L-1} - {\xi}^{i,*}_L \Vert_{P_i}^2 + 2\sqrt{V_t^*} \Vert \hat{\xi}^{i}_{L-1} -{\xi}^{i,*}_L \Vert_{P_i}
        )  
        .
    $
    From Assumption~\ref{asm:extended_state_bound}, and with $\sqrt{V_t^*} \le V_t^* + 1$
    \begin{align}\label{eq:IO_decrease}
        &V_{t+1}^* - V_t^* \le {\sum_{i=1}^M} ((L{-1}) (\bar{\sigma}(\sigma_i^{\prime}(\Omega_i)) + 2V_t^* \sqrt{\sigma_i^{\prime}(\Omega_i)})) + \mu 
        \notag
        \\
        &\phantom{\le{}} - \Vert y_t \Vert_{Q}^2 - \Vert u_t \Vert_{R}^2 + \sum_{i=1}^M \bar{\sigma}(\tilde{\sigma}_i(\Omega_i))
        + 2V_t^* \sqrt{\tilde{\sigma}_i(\Omega_i)}
    \end{align}
    with $\bar{\sigma}(r) = r + 2\sqrt{r}$.
    As in the proof of~\cite[Theorem 8]{berberich2021on}, since $(A, C)$ is observable by Assumption~\ref{asm:ctrb_pe_IO}, $(\tilde{A}, \tilde{C})$ is detectable and~\eqref{eq:extended_state_global_system} admits an input-output-to-state stability Lyapunov function $W(\xi)  = \Vert \xi \Vert_{P_W}^2$ satisfying
    \begin{equation}\label{eq:IOSS_decrease}
        W(\tilde{A}\xi + \tilde{B}u) - W(\xi) \le -\frac{1}{2}\Vert \xi \Vert^2 + c_1 \Vert y \Vert_2^2 + c_2 \Vert u \Vert_2^2,        
    \end{equation}
    with $c_1, c_2 > 0$ and $P_W \succ 0$, for all $u \in \mathbb{R}^m$, $\xi \in \mathbb{R}^{n_\xi}$ and $y = \tilde{C}\xi + \tilde{D}u$ (cf.~\cite[]{cai2008input}).
    Consider now $\mathcal{W}_t = \gamma W(\xi_t) + V_t^*$ with $\gamma = \frac{\lambda_{\min}(Q, R)}{\max(c_1, c_2)}$.
    Clearly, $\gamma\Vert\xi_t \Vert_{P_W}^2 \le \mathcal{W}_t \le c_{\mathrm{ub}}\Vert \xi_t \Vert^2 + \gamma \Vert \xi_t \Vert_{P_W}^2$,
    where the lower bound follows from $V_t^* \ge 0$ and the upper bound from Assumption~\ref{asm:V_t_upper_bound}.
    Hence, from Assumption~\ref{asm:V_t_upper_bound} and combining~\eqref{eq:IO_decrease} with~\eqref{eq:IOSS_decrease},
    $
    \mathcal{W}_{t+1} - \mathcal{W}_{t} 
    \le -\frac{\gamma}{2} \Vert \xi_t \Vert^2 
    + {\sum_{i=1}^M} 
    (
        (L-1) 
        \bar{\sigma}(\sigma_i^{\prime}(\Omega_i)) + \bar{\sigma}(\tilde{\sigma}_i(\Omega_i))
    )
    + 2 c_{\mathrm{ub}} \Vert \xi_t \Vert^2 
    \sum_{i=1}^M 
    (
        (L-1)   \sqrt{\sigma_i^{\prime}(\Omega_i)} + \sqrt{\tilde{\sigma}_i(\Omega_i)} 
    )
    + \mu.
    $
    Thus, there exist $\Omega_i$ sufficiently small such that
    $
    \mathcal{W}_{t+1} - \mathcal{W}_{t} {\le} -\tilde{\gamma} \Vert \xi_t \Vert^2 
    + {\sum_{i=1}^M} (L-1) \bar{\sigma}(\sigma_i^{\prime}(\Omega_i))
    + \bar{\sigma}(\tilde{\sigma}_i(\Omega_i)) + \mu
    $
    for some $\tilde{\gamma} > 0$.
    Hence, there also exists $\sigma \in \mathcal{K}_\infty$ such that
    $
        \mathcal{W}_{t+1} - \mathcal{W}_{t} \le  -\tilde{\gamma} \Vert \xi_t \Vert^2 + \sigma ( \Omega ) + \mu,
    $
    and the claim follows~\cite[Theorem 2.4]{Gruene2014}.
    \hspace{\fill} \qed
\end{pf}

\section{Numerical example}
We consider the same example as in~\cite[]{alonso2021data}: a system comprising a chain of 64 subsystems with dynamics
\begin{subequations}\label{eq:example_sys}
    \begin{align}
        x_{t+1}^i &=
        \begin{bmatrix}
            1 & 0.2 \\ -\frac{k_i}{5m_i} & 1 - \frac{d_i}{5m_i}
        \end{bmatrix}x_t^i
        + \begin{bmatrix}
            0 \\  u_t^i
        \end{bmatrix}
        + {\sum_{j\in\mathcal{N}_i}} \begin{bmatrix}
            0 \\ k_{ij}y_t^j
        \end{bmatrix},
        \\
        y_{t}^i &= \begin{bmatrix}
            \frac{0.2}{m_i} & 0
        \end{bmatrix}x_t^i.
    \end{align}
\end{subequations}
The parameters are set to $m_i = 1$, $d_i = 0.75$, $k_{ij} = 1.25$ and $k_i = \sum_{j\in\mathcal{N}_i} k_{ij}$ for all $i \in \mathcal{V}$ and $j \in \mathcal{N}_i$.
We consider input-output data of each subsystem of length $N = 100$.
We choose $L = 5$, $Q_i = R_i = I$, $U_i = [-2,2]$ and design terminal ingredients as discussed with $\epsilon_i = 10^{-5}$ for all $i \in \mathcal{V}$. 
For the consistency constraint, we choose $\Omega_i = 0.01$ for all $i \in \mathcal{V}$.
We use a suboptimal feasible trajectory as the initially feasible trajectory.
The LMIs in Lemma~\ref{lm:terminal_ingredients_lmi} were solved using YALMIP~\cite[]{lofberg2004yalmip} and MOSEK~\cite[]{mosek.2020}, whereas MOSEK was used for~\eqref{eq:D3MPC_IO}.
We plot the closed-loop output evolution of subsystems $i\in\mathbb{I}_{[1,7]}$ in Figure~\ref{fig:states}, displaying (practical) convergence, as expected from Theorem~\ref{thm:stability}.
\setlength\fheight{5.2cm}
\setlength\fwidth{0.85\columnwidth}
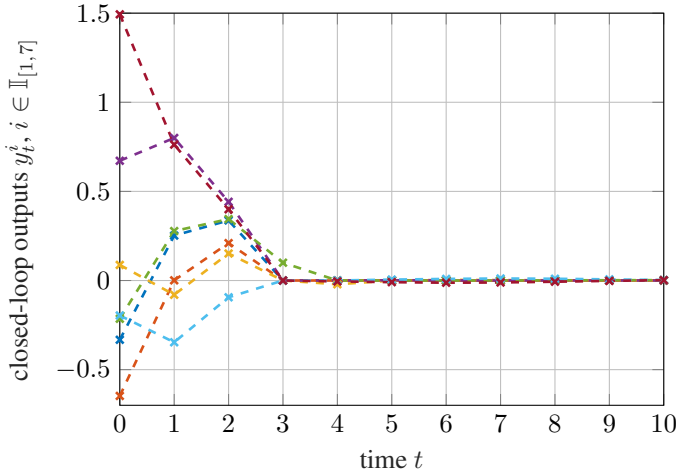
\begin{figure}[t]
    \begin{center}
        \definecolor{mycolor1}{rgb}{0.00000,0.44700,0.74100}%
\definecolor{mycolor2}{rgb}{0.85000,0.32500,0.09800}%
\definecolor{mycolor3}{rgb}{0.92900,0.69400,0.12500}%
\definecolor{mycolor4}{rgb}{0.49400,0.18400,0.55600}%
\definecolor{mycolor5}{rgb}{0.46600,0.67400,0.18800}%
\definecolor{mycolor6}{rgb}{0.30100,0.74500,0.93300}%
\definecolor{mycolor7}{rgb}{0.63500,0.07800,0.18400}%
\begin{tikzpicture}

\begin{axis}[%
width=0.951\fwidth,
height=\fheight,
at={(0\fwidth,0\fheight)},
scale only axis,
xmin=0,
xmax=10,
xlabel style={font=\color{white!15!black}},
xlabel={time $t$},
ymin=-0.7,
ymax=1.5,
ylabel style={font=\color{white!15!black}},
ylabel={closed-loop outputs $y_t^i$, $i \in \mathbb{I}_{[1,7]}$},
axis background/.style={fill=white},
xtick = {0,1,2,3,4,5,6,7,8,9,10},
xmajorgrids,
ymajorgrids
]
\addplot [color=mycolor1, dashed, line width=1.0pt, mark=x, mark options={solid, mycolor1}]
  table[row sep=crcr]{%
0	-0.332417881655132\\
1	0.252284183906497\\
2	0.337126435400377\\
3	-4.55583917258195e-07\\
4	-1.39075909899589e-07\\
5	1.90670097488521e-06\\
6	-3.88039201126844e-06\\
7	5.19008907140517e-06\\
8	1.37461712910181e-05\\
9	1.42932409445962e-05\\
10	9.10381699092966e-06\\
};

\addplot [color=mycolor2, dashed, line width=1.0pt, mark=x, mark options={solid, mycolor2}]
  table[row sep=crcr]{%
0	-0.647415253741755\\
1	0.00113230746911874\\
2	0.209841441790759\\
3	-7.46944898821766e-07\\
4	-5.74183601997902e-07\\
5	0.000393747584498172\\
6	5.19710394790707e-05\\
7	-0.000144421598236022\\
8	-0.00014226602802303\\
9	-7.27796138146886e-05\\
10	-7.66625798220844e-06\\
};

\addplot [color=mycolor3, dashed, line width=1.0pt, mark=x, mark options={solid, mycolor3}]
  table[row sep=crcr]{%
0	0.0876028926220442\\
1	-0.0798792534601134\\
2	0.151642110143023\\
3	-1.25854701771289e-06\\
4	-0.0210326700094772\\
5	-0.000412572242904474\\
6	0.00355188981131072\\
7	0.00261741590717257\\
8	0.00124199791906143\\
9	0.000417226333123466\\
10	7.40283860594104e-05\\
};

\addplot [color=mycolor4, dashed, line width=1.0pt, mark=x, mark options={solid, mycolor4}]
  table[row sep=crcr]{%
0	0.671884690287003\\
1	0.799842706782956\\
2	0.441801755119533\\
3	-7.01069177644342e-06\\
4	-2.52289684254947e-06\\
5	0.00140685395743567\\
6	0.000859593916919366\\
7	0.000521477952965199\\
8	0.000421149151769384\\
9	0.000291760576180522\\
10	0.000297685496583711\\
};

\addplot [color=mycolor5, dashed, line width=1.0pt, mark=x, mark options={solid, mycolor5}]
  table[row sep=crcr]{%
0	-0.214528539343632\\
1	0.277487465338728\\
2	0.344678714219835\\
3	0.0999934595703973\\
4	0.000174675017030168\\
5	-1.83047744837239e-05\\
6	-3.59812104822055e-05\\
7	0.000288689426856337\\
8	0.00106525043517358\\
9	0.00160385372863914\\
10	0.00213826681039109\\
};

\addplot [color=mycolor6, dashed, line width=1.0pt, mark=x, mark options={solid, mycolor6}]
  table[row sep=crcr]{%
0	-0.196088598580527\\
1	-0.346472425998677\\
2	-0.094093363063477\\
3	0.000172016734959257\\
4	0.00019807680009265\\
5	0.00506082399948227\\
6	0.00904201195649001\\
7	0.011463527190406\\
8	0.00987704144735169\\
9	0.00639563562363321\\
10	0.00277842155377428\\
};

\addplot [color=mycolor7, dashed, line width=1.0pt, mark=x, mark options={solid, mycolor7}]
  table[row sep=crcr]{%
0	1.49324233345504\\
1	0.762618323537696\\
2	0.399533403538971\\
3	3.13416073467465e-05\\
4	-0.0049924131386252\\
5	-0.00924038217266165\\
6	-0.0123364209433152\\
7	-0.0109637712336443\\
8	-0.00676248504406018\\
9	-0.00187842020915951\\
10	0.00220413275833486\\
};

\end{axis}
\end{tikzpicture}%
        \caption{Evolution of the closed-loop outputs of~\eqref{eq:example_sys} for $i\in\mathbb{I}_{[1,7]}.$}
        \label{fig:states}
    \end{center}
\end{figure}

\section{Conclusion}
We have proposed a direct data-driven distributed MPC scheme for a group of dynamically coupled LTI systems.
Each local MPC optimisation problem uses only past measured input-output data for the prediction, without any prior system identification step.
We showed that if the dynamic coupling is sufficiently weak, the data-driven distributed MPC scheme is recursively feasible and practically stabilises the origin of the global system.
The main mechanism are so-called consistency constraints, i.e. keeping close to a previously communicated trajectory, based on the model-based approach in~\cite[]{Dunbar.2007}.
This enables a non-iterative parallel distributed MPC scheme with minimal communication.
The complexity of the scheme does not increase with the total number of subsystems, but only with the number of neighbours of each subsystem.
Future research will investigate deriving suitable bounds for the consistency based on input-output data, as well as data-driven design of terminal ingredients that take the dynamic coupling into account.

{\bibliography{Literature}}

\end{document}